%% file: main.tex
\documentclass[conference]{IEEEtran}

\usepackage[
  pass,
]{geometry}

\usepackage[latin1]{inputenc}
\usepackage{amsmath,bbm,subfigure}

\usepackage{graphicx}
\usepackage{acronym}
\usepackage{changes}
\usepackage{lipsum}
\definechangesauthor[name={Authors}, color=blue]{authors}

\usepackage{acronym}

\usepackage[acronym]{glossaries}
\usepackage{glossary-mcols}
\usepackage{url,comment}
\usepackage{booktabs}
\usepackage{flushend}

\usepackage{amsmath}
\usepackage{algorithm}
\usepackage[noend]{algpseudocode}
\usepackage{siunitx} 

\usepackage{array}
\newcolumntype{L}[1]{>{\raggedright\let\newline\\\arraybackslash\hspace{0pt}}m{#1}}
\newcolumntype{C}[1]{>{\centering\let\newline\\\arraybackslash\hspace{0pt}}m{#1}}
\newcolumntype{R}[1]{>{\raggedleft\let\newline\\\arraybackslash\hspace{0pt}}m{#1}}

\makeatletter
\def\BState{\State\hskip-\ALG@thistlm}
\makeatother

\input{acronyms}

\newglossarystyle{modsuper}{%
  \glossarystyle{super}%
}

\begin{document}
\title{Cloud-RAN Factory: Instantiating virtualized mobile networks with ONAP}


\author{\IEEEauthorblockN{Veronica Quintuna Rodriguez, Romuald Corbel,  Fabrice Guillemin,  Alexandre Ferrieux}\\ 
\IEEEauthorblockA{Orange Labs, 2 Avenue Pierre Marzin,  22300 Lannion, France} \{veronica.quintunarodriguez, romuald.corbel, fabrice.guillemin, alexandre.ferrieux\}@orange.com}

\maketitle  
\begin{abstract}
In this demo, we exhibit the negotiation based on the TM Forum framework (Customer Facing Service and Resource Facing Service) and the deployment of a fully virtualized end-to-end mobile network (including a RAN desegregated into Remote Unit, Distributed Unit and Centralized Unit) by using ONAP, an open-source network automation platform. The various components of mobile network are containerized and deployed by ONAP on top of Kubernetes. The demo is the first illustration of an end-to-end mobile network, which is fully virtualized up to the remote unit, whose architecture is compatible with the Open-RAN framework, and which implements a PHY layer on the basis of 3GPP 7.3 functional split in Open Air Interface code.

\end{abstract}


\section{Introduction}

In the context of telecommunication networks, virtualization techniques have enabled the emergence of \gls{NFV}, which radically changes the architecture and the operation of  networks \cite{ETSI_NFVArch,virtualization}.  \gls{NFV} greatly impacts specific data plane functions such as Firewalls, Deep Packet Inspection, etc.  but also  core control functions of mobile networks, for instance via the development of  \gls{vEPC} in 4G and boosting the design of a completely new control architecture in 5G  (Service Based Architecture, SBA). NFV now also applies to more latency sensitive functions such as those of \gls{RAN}. The development of Software Defined Radio (SDR) is long lasting task and the 4G \gls{OAI} software suite (including \gls{RAN} and core network) is today sufficiently stable to run field trials in the context of \gls{NFV} and offers a development platform to investigate new features of \gls{RAN}  \cite{OAI}; the development of an 5G \gls{OAI} is also rapidly progressing.

The emergence of \gls{NFV} and its application to core and \gls{RAN} has boosted the development of various open source communities as \gls{ORAN}, \gls{ONAP}, \gls{OCN}, among others. The automation of the deployment of \glspl{VNF} and their life-cycle management is addressed by the \gls{ONAP} community.  The \gls{ONAP} platform is  roughly based on two main phases: the design time and the run time. Each phase is composed of a number of modules in charge of performing specific tasks.  The onboarding and the deployment of a \gls{vEPC} using ONAP has notably been addressed in \cite{icin2020} in the context of network slicing. 

\gls{ORAN} community \cite{oranWebSite} addresses more specifically the \gls{RAN} by specifying the basic building blocks of an open RAN, the interfaces between these  blocks as well as the methods of splitting \gls{RAN} functions. \gls{ORAN} also introduces various controllers (e.g., \gls{RIC}) to optimize the \gls{RAN} performance. In this paper, we shall focus on functional blocks along the data path: \gls{RU}, \gls{DU} and \gls{CU}. This various modules execute the RAN functions: PDCP, RLC, MAC, and  PHY functions. \gls{ORAN} also considers the orchestration of \gls{RAN} functions; either in cooperation with ONAP or via a standalone orchestration platform.

It is worth nothing that the desegregation and the softwarization of RAN functions have been used to define open architectures and to implement cloud-native RAN functions under the concept of \gls{C-RAN} or Cloud-RAN, see for instance \cite{chinaMobile,jsac}. 

The cloudification of RAN functions has first considered "high" functions (PDCP and a part of RLC functions  to centrally manage the  X2 interface). Those functions are usually implemented in the \gls{CU} functional block.

In \cite{jsac,contribcran}, the virtualization of PHY functions (notably the channel encoding and decoding function) has been implemented by taking benefit of the parallelization by means of thread-pool mechanisms. An implementation in \gls{OAI} RAN code has shown that  it is possible to significantly reduce the execution time of coding/decoding  and to increase the distance between the intra-PHY functions (channel coding and MAC functions on the one hand and the rest of PHY functions on the other hand). Channel Coding and MAC functions are now executed  in the \gls{DU} functional block. \gls{CU} and \gls{DU} functional blocks can be implemented in Virtual Machines managed by OpenStack or Containers managed by Kubernetes.

In this demo, we go one step further by proposing to virtualize the \gls{RU} functional block. The rationale motivating this approach is that  physical infrastructures hosting hardware to transmit and receive radio signals (antenna, power supply, etc.) are  more and more equipped with mini data centers, which can virtualized. In that case, an operator can push its own RU software into such data centers and the physical site can  then be shared by several operators. This is perfectly in line with the  rapidly emerging market  of TowerCos and with the need for rationalizing the installation of infrastructures hosting antennas  in order to reduce CAPEX.

The proposed demo shows how a fully virtualized technical chain corresponding to a  mobile network (including RAN and core control) can be instantiated by ONAP on a multi-cloud platform, i.e., supporting Kubernetes and Openstack as \glspl{VIM}. 

The organization of this demo paper is as follows: In Section~\ref{elements}, we introduce the various elements of the demo and in Section~\ref{show} we  describe what is shown by the demo. Some concluding remarks are presented in Section~\ref{conclusion}. 

\section{Elements of the demo}
\label{elements}

\begin{figure}[hbtp]
\centering \scalebox{.4}{\includegraphics{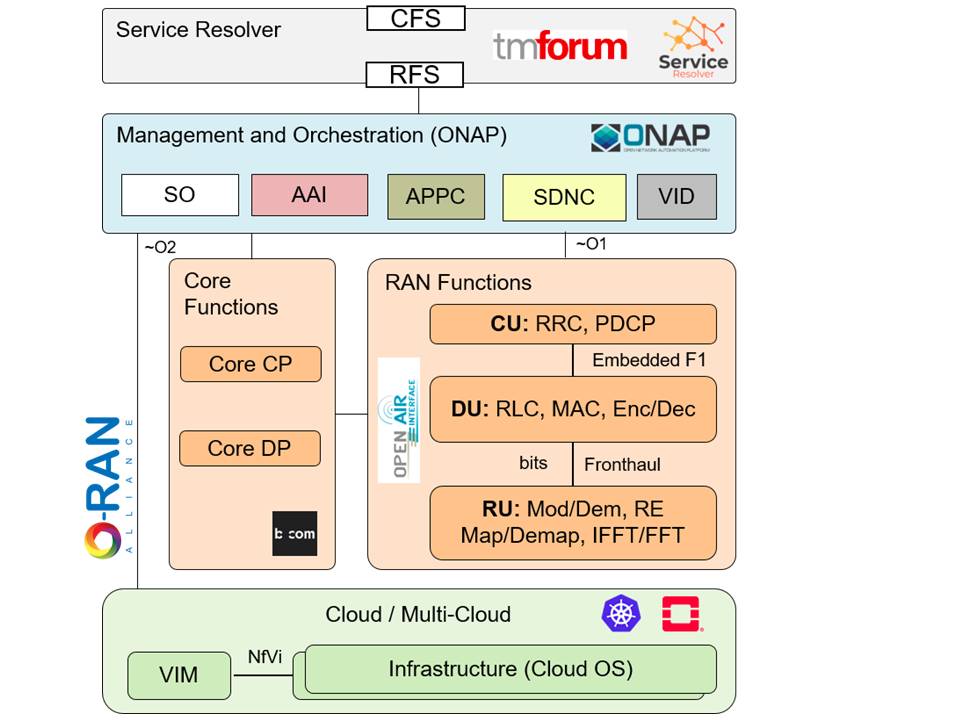}}
\caption{Elements of the demo according to O-RAN and ODA architectures.\label{fig:elements}}
\end{figure}

The objective of the demo is to show how a fully virtual mobile technical chain (including the core control network and the \gls{RAN}) can be negotiated between a customer and the network and automatically instantiated by \gls{ONAP}. The various elements involved in the demo are presented in Figure~\ref{fig:elements}. The negotiation is performed through the \gls{CFS}.

The technical chain is called a network service in the sense of \gls{ONAP}; it is worth noting that such a network service can itself be composed of several network services (e.g., one for the control plane, another one for the core data plane and a last one for the \gls{RAN}). The negotiation between the customer and the network is performed through the \gls{CFS} as specified by the TM Forum \cite{tmf641}. The service template chosen via the \gls{CFS} is then translated by the Service Resolver into a resource request via the \gls{RFS}, this latter request  is then passed to the ONAP platform.

The \gls{ONAP} platform embeds all the virtualized components of the desired network services. These various components have been previously onboarded during the design time in the form of VMs for the core and containers for the RAN network functions.

\gls{ONAP} is composed of several modules where the most relevant in this demo are the Service Orchestrator (SO) to deploy the various CNFs, the SDN Controller to deals with the various VNFs and the \gls{AAI}  that provides real-time views of available Resources and Services and their relationships. Service resolver consumes AAI information by means of the \gls{NBI} module. 

The cloud platform hosting the VNFs/CNFs is based on OpenStack/Kubernetes. The interface between the orchestrator and the cloud platform is similar to the O2 interface specified by \gls{ORAN} and the one between the orchestrator and the RAN functions to the O1 interface (see \cite{ORANWG4-CUS}).

\begin{figure}[hbtp]
\centering
  \includegraphics[scale=0.46, trim=0 180 170 0, clip] {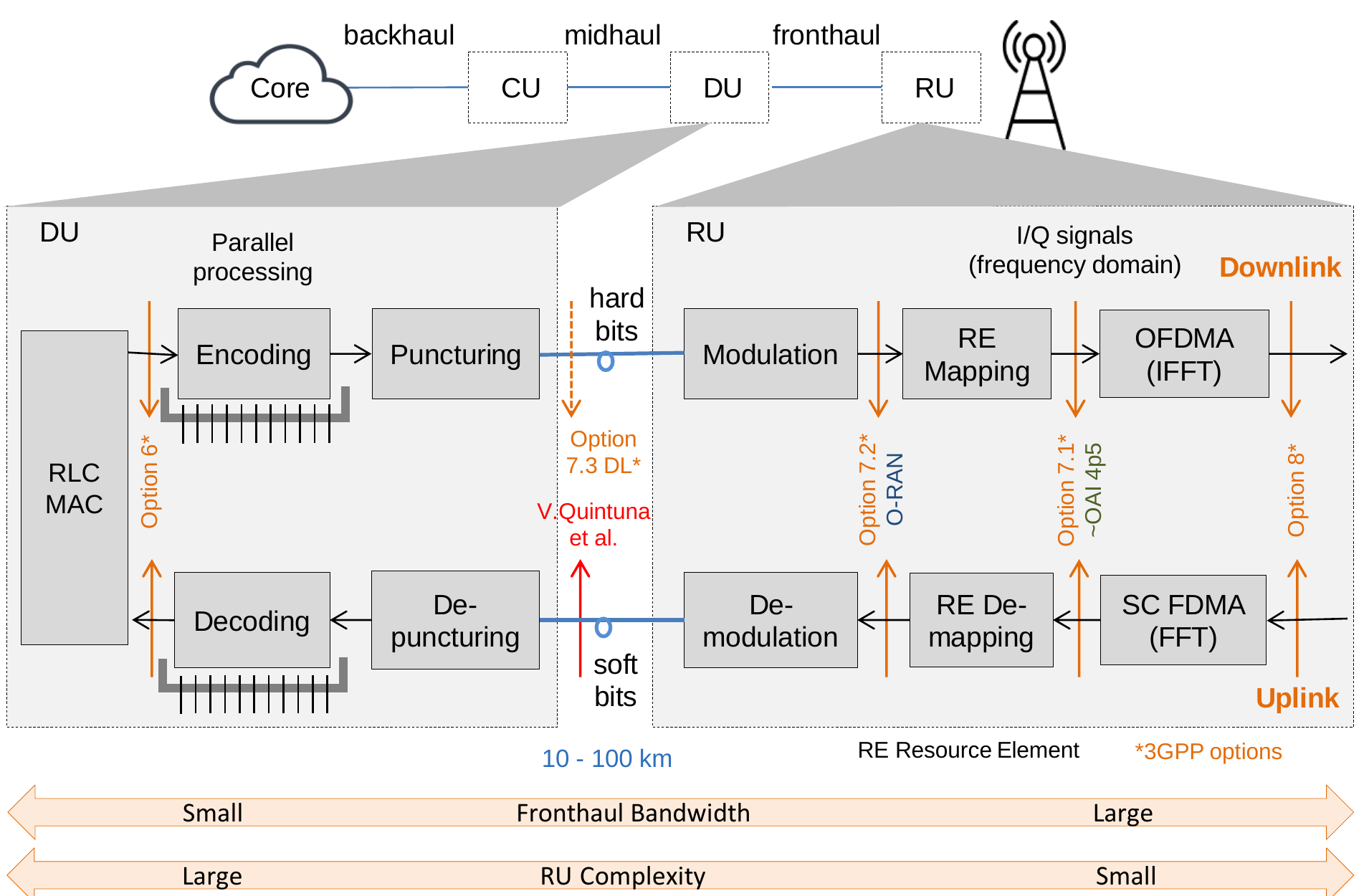}
    \caption{Implemented functional split architecture.}
    \label{fig:fs73}
\end{figure}

The functions contained in the \gls{RU}, \gls{DU} and the \gls{CU} are based on OAI code. While ORAN is based on the  7.2 functional split (in the terminology of 3GPP \cite{3GPP38_801}), for separating the low PHY functions, we have here developed, as detailed in \cite{iwcmc2020}, the 7.3 split in OAI code. This split consists of transmitting hard bits between the encoder (after puncturing) and Modulation functions in the downlink instead of compressed I/Q signals, and soft bits in the uplink. It is worth noting that in the 3GPP specifications \cite{ORANWG4-CUS}, the 7.3 functional split was only considered in the downlink. In the implementation presented in \cite{iwcmc2020}, it applies in both  downlink and uplink, as illustrated in Figure~\ref{fig:fs73}.

The 7.3 functional split has been developed because it requires less fronthaul capacity (bandwidth between \gls{RU} and \gls{DU}) than 7.2 split. Moreover, the parallelization of coding/decoding code blocks significantly reduces the execution of these tasks, which in turn allows a larger distance between \gls{RU} and \gls{DU}.

In this demo the \gls{DU} and \gls{CU} have been collocated and deployed as a single container, hence the F1 interface is embedded. Globally, the demo implements on the basis of OAI code a mobile technical chain, which is  compatible with the ORAN architecture for the data plane. The various containers executing the RAN and control plane functions are instantiated by means of ONAP. The interfaces between the cloud platform and the SO as well as that between SDNC and the VNFs are similar to the O1 and O2 interfaces introduced by ORAN, respectively. Thus, the demo is a first illustration of an ORAN compatible end-to-end mobile network orchestrated by ONAP.

\section{What is shown by the demo}
\label{show}

\begin{figure}[hbtp]
\centering
 \scalebox{.35}{\includegraphics {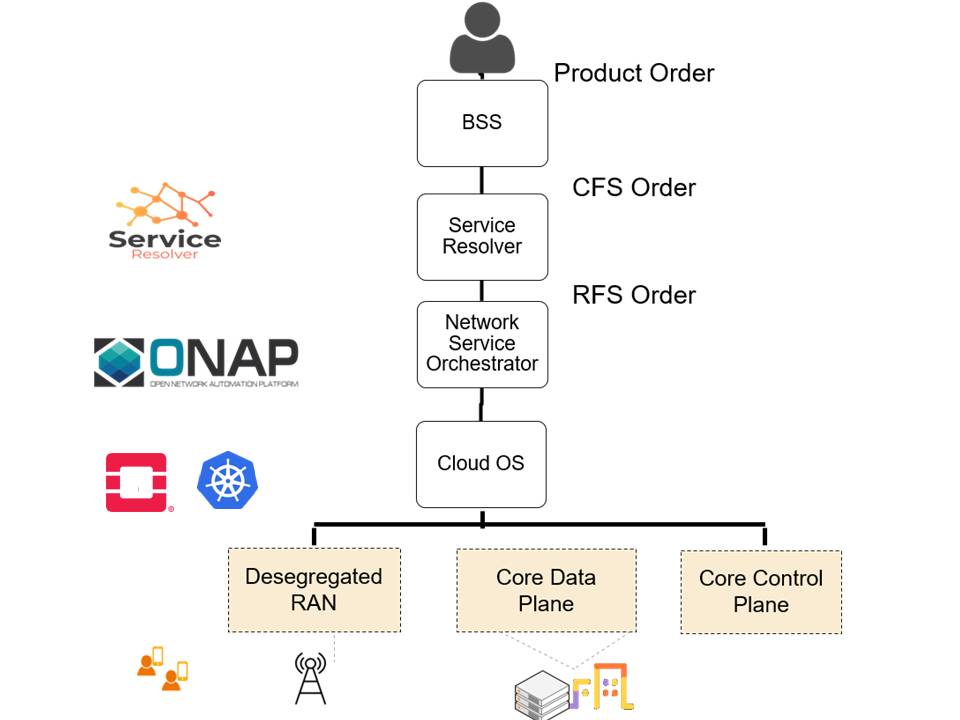}}
    \caption{Cloud-RAN from negotiation to instantiation.}
    \label{fig:global}
\end{figure}

The demo realizes the global picture given by Figure~\ref{fig:global}. The customer negotiates with the network the setup of a virtual end-to-end mobile network tailored to some specific requirements (number of customers, bandwidth, etc.). The request is then passed  via the CFS  to the Service Resolver, which subsequently translates it into a resource request via the RFS. The resource request in terms of VNFs/CNFs (with associated  CPU and storage requirements), bandwidth and connectivity (including the radio coverage)  is then handled by ONAP, which deploys the VNFs of the core control and data planes on a Openstack tenant as well as the containerized \gls{RU}, \gls{DU} and \gls{DU} units on a Kubernetes cluster, this latter formed by a remote node hosting the RU. For the core control plane, we have used the \gls{WEF} developed by b$<>$com; this a completely virtualized core network (full 4G and preliminary 5G); see \cite{icin2020} for details in the context of network slicing.

\begin{figure}[hbtp]
\centering
 \scalebox{.36}{\includegraphics {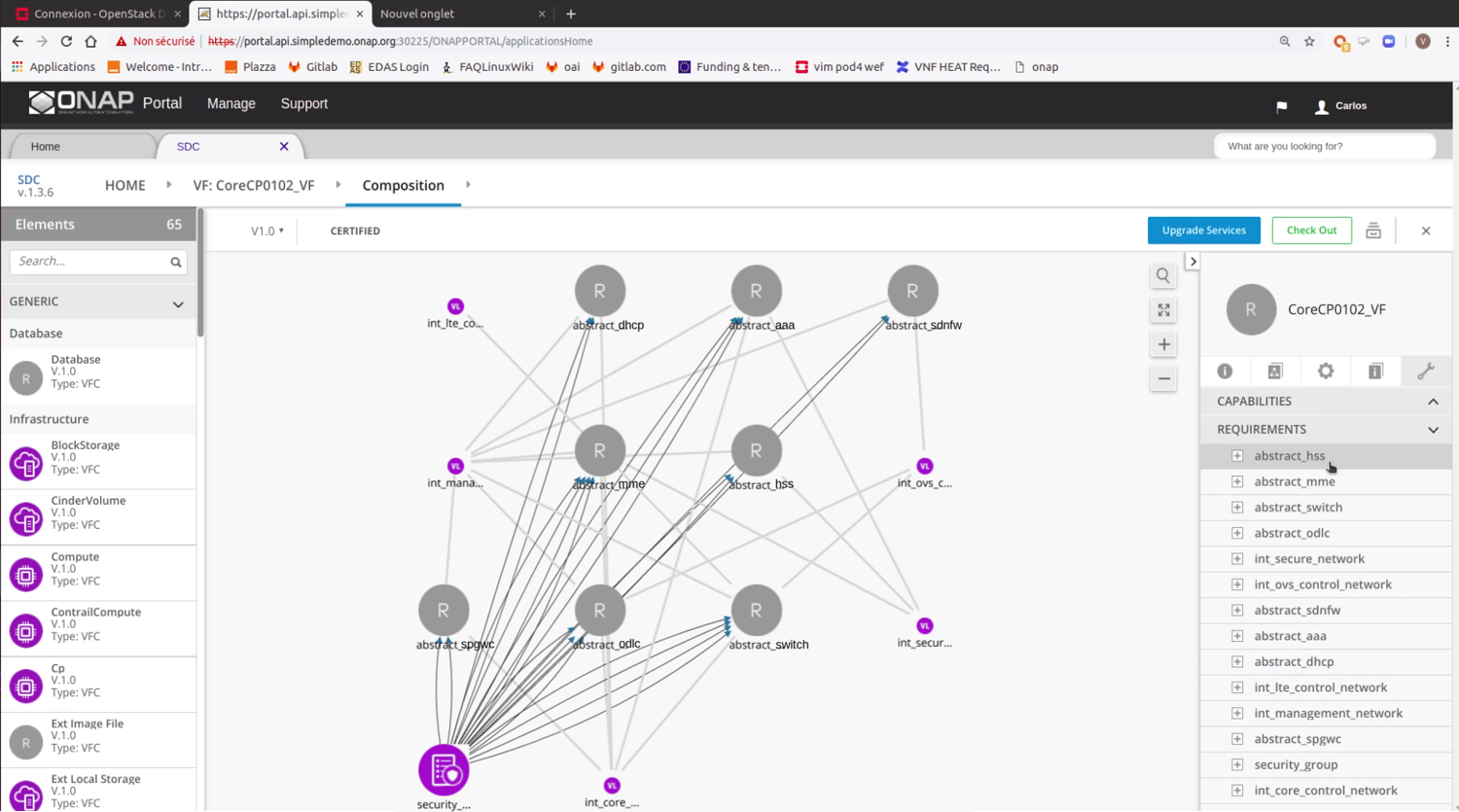}}
    \caption{ONAP GUI showing the deployment of the 4G WEF components.}
    \label{fig:onapgui}
\end{figure}

The ONAP GUI is used to verify the deployment of the various VNFs and the associated interconnection network. Figure~\ref{fig:onapgui} is a snapshot of the OANP GUI showing the deployment of the elements of the 4G control plane of the WEF and their internal interconnections.

\begin{figure}[hbtp]
\centering
 \scalebox{.39}{\includegraphics {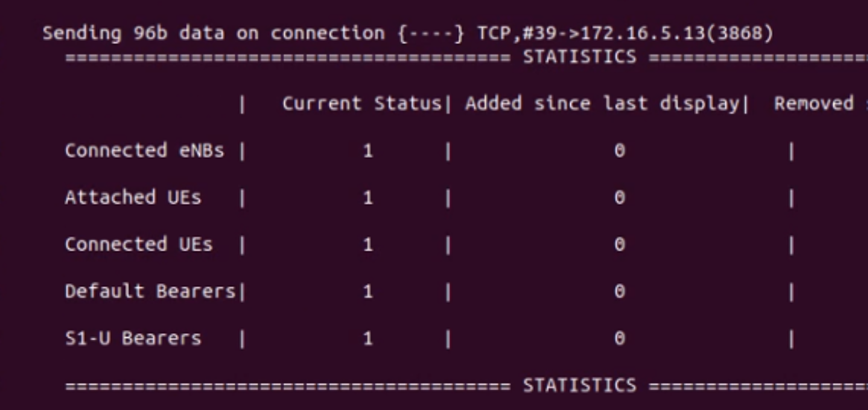}}
    \caption{Monitoring interface of the OAI MME VM.}
    \label{fig:mmegui}
\end{figure}

Once the end-to-end virtualized mobile network is deployed, we verify that the connectivity is really offered by attaching a commercial UE. Figure~\ref{fig:mmegui} illustrates the monitoring interface provided by the OAI software, namely the monitoring of the \gls{MME} which shows the successful connectivity.

\section{Conclusion}
\label{conclusion}
The demo illustrates the negotiation and the instantiation of an end-to-end mobile network by using the TMF specifications for the negotiation and ONAP for the deployment. The RAN (involving the RU, DU and CU) is entirely containerized and deployed by using K8s. The used PHY layer is the 7.3 split implemented in OAI code. The demo is the first realization of a fully end-to-end virtual mobile network compatible with the ORAN architecture and deployed by ONAP.

\bibliographystyle{IEEEtran}
\bibliography{biblo}
\flushend

\end{document}

%% file: acronyms.tex
\newacronym{AAA}{AAA}{Authentication, Authorization, and Accounting}
\newacronym{RSC}{RSC}{Recursive  Systematic Convolutional}
\newacronym{LLR}{LLR}{Log-Likelihood Ratio}
\newacronym{FS}{FS}{Functional Split}
\newacronym{BBU}{BBU}{Base Band Unit}
\newacronym{COTS}{COTS}{Commercial off-the-shelf }
\newacronym{VNF}{VNF}{Virtualized Network Function}
\newacronym{VNF FG}{VNF FG}{VNF Forwarding Graph}
\newacronym{NFV}{NFV}{Network Function Virtualization}
\newacronym{GPP}{GPP}{General Purpose Processor}
\newacronym{vEPC}{vEPC}{virtual Evolved Packet Core}
\newacronym{LTE}{LTE}{Long Term Evolution}
\newacronym{uRLLC}{uRLLC}{Ultra-Reliable Low-Latency Communications}
\newacronym{eMBB}{eMBB}{enhanced Mobile BroadBand}
\newacronym{mMTC}{mMTC}{massive Machine Type Communications}
\newacronym{OSM}{OSM}{Open Source MANO}
\newacronym{C-EPC}{C-EPC}{Cloud-EPC}
\newacronym{EPCaaS}{EPCaaS}{EPC as a Service}
\newacronym{TDD}{TDD}{Time Division Duplex}
\newacronym{UE}{UE}{User Equipment}
\newacronym{HARQ}{HARQ}{Hybrid Automatic Repeat-Request}
\newacronym{PRB}{PRB}{Physical Resource Blocks}
\newacronym{MCS}{MCS}{Modulation and Coding Scheme}
\newacronym{CQI}{CQI}{Channel Quality Indicator}
\newacronym{DC}{DC}{Dedicated Core}
\newacronym{RR}{RR}{Round Robin}
\newacronym{G}{G}{Greedy}
\newacronym{VPN}{VPN}{Virtual Private Network}
\newacronym{MPLS}{MPLS}{Multiprotocol Label Switching}
\newacronym{OWL}{OWL}{Web Ontology Language}
\newacronym{NST}{NST}{Network Slice Template}
\newacronym{NSST}{NSST}{Network Slice Subnet Template}
\newacronym{NSMF}{NSMF}{Network Slice Management Function}
\newacronym{NSSMF}{NSSMF}{Network Slice Subnet Management Function}
\newacronym{CSMF}{CSMF}{Communication Service Management Function}
\newacronym{FCAPS}{FCAPS}{Fault-management Configuration Accounting Performance and Security}
\newacronym{RF}{RF}{Radio Frequency}
\newacronym{RIC}{RIC}{RAN Intelligent Controller}

\newacronym{CNF}{CNF}{Cloud-Native Network Function}
\newacronym{PLMN}{PLMN}{Public Land Mobile Network}
\newacronym{CLAMP}{CLAMP}{Closed Loop Automation Management Platform}

\newacronym{FDD}{FDD}{Frequency Division Duplex}
\newacronym{OFDM}{OFDM}{Orthogonal Frequency Division Multiplexing}

\newacronym{VM}{VM}{Virtual Machine}
\newacronym{PDCP}{PDCP}{Packet Data Convergence Protocol}
\newacronym{MAC}{MAC}{Medium Access Control}
\newacronym{RLC}{RLC}{Radio Link Control}
\newacronym{RRC}{RRC}{Radio Resource Control}
\newacronym{AM}{AM}{Acknowledged Mode}
\newacronym{UM}{UM}{Unacknowledged Mode}
\newacronym{TM}{TM}{Transparent Mode}
\newacronym{MIMO}{MIMO}{Multiple Input Multiple Output}
\newacronym{MISO}{MISO}{Multiple Input Single Output}
\newacronym{SIMO}{SIMO}{Single Input Multiple Output}
\newacronym{SISO}{SISO}{Single Input Single Output}
\newacronym{MCC}{MCC}{Mobile Country Code}
\newacronym{MNC}{MNC}{Mobile Network Code}
\newacronym{S-TMSI}{S-TMSI}{Shortened Temporary Mobile Subscriber Identity}
\newacronym{IMSI}{IMSI}{International Mobile Subscriber Identity}
\newacronym{DRB}{DRB}{Dedicated Radio Bearer}
\newacronym{GUMMEI}{GUMMEI}{Globally Unique MME Identity}

\newacronym{PCI}{PCI}{Physical-layer Cell Identity}
\newacronym{ROHC}{ROHC}{Robust Header Compression}
\newacronym{SN}{SN}{Sequence Number}
\newacronym{RAR}{RAR}{Random Access Response}
\newacronym{C-RNTI}{C-RNTI}{Cell Radio Network Temporary Identifier}
\newacronym{BSR}{BSR}{Buffer Status Report}
\newacronym{DRX}{DRX}{Discontinuous Reception}
\newacronym{PHR}{PHR}{Power Head Room}
\newacronym{PUSCH}{PUSCH}{Physical Uplink Shared Channel}
\newacronym{ADM}{ADM}{Activation/Deactivation MAC}
\newacronym{GP}{GP}{Gap Period}

\newacronym{RE}{RE}{Resource Element}
\newacronym{RB}{RB}{Resource Block}
\newacronym{REG}{REG}{Resource Element Group}
\newacronym{CSRS}{CSRS}{Cell-Specific Reference Signal}
\newacronym{IFFT}{IFFT}{Inverse Fast Fourier Transform}
\newacronym{OFDMA}{OFDMA}{Orthogonal Frequency Division Multimple Access}
\newacronym{CRC}{CRC}{Cyclic Redundancy Check}
\newacronym{SFC}{SFC}{Service Function Chain}

\newacronym{eNB}{eNB}{Evolved NodeB}
\newacronym{RAN}{RAN}{Radio Access Network}
\newacronym{ARQ}{ARQ}{Automatic Repeat reQuest}
\newacronym{NAS}{NAS}{Non-Access Stratum}
\newacronym{MME}{MME}{Mobility Management Entity}
\newacronym{MIB}{MIB}{Master Information Block}
\newacronym{SIB}{SIB}{System Information Block}
\newacronym{RSRP}{RSRP}{Reference Signal Received Power}
\newacronym{RAT}{RAT}{Radio Access Technologie}
\newacronym{ACK}{ACK}{Acknowledge}
\newacronym{NACK}{NACK}{Negative acknowledge}
\newacronym{PDCCH}{PDCCH}{Physical Downlink Control Channel}
\newacronym{SAW}{SAW}{Stop and Wait}
\newacronym{TTI}{TTI}{Transmission Time Interval}
\newacronym{RRH}{RRH}{Radio Remote Head}
\newacronym{SNIR}{SNIR}{Signal-to-Noise-plus-Interference Ratio}
\newacronym{WCET}{WCET}{Worst Case Execution Time}
\newacronym{GPC}{GPC}{General Purpose Computer}
\newacronym{KPI}{KPI}{Key Performance Indicator}
\newacronym{OAI}{OAI}{Open Air Interface}
\newacronym{IMS}{IMS}{IP Multimedia Subsystem}
\newacronym{vIMS}{vIMS}{virtual IP Multimedia Subsystem}
\newacronym{EPC}{EPC}{Evolved Packet Core}
\newacronym{SDN}{SDN}{Software Defined Network}
\newacronym{C-RAN}{C-RAN}{Cloud-RAN}
\newacronym{OS}{OS}{Operating System}
\newacronym{TB}{TB}{Transport Block}
\newacronym{TBS}{TBS}{Transport Block Size}
\newacronym{QCI}{QCI}{QoS Channel Indicator}

\newacronym{GPU}{GPU}{Graphics Processing Unit}
\newacronym{CPU}{CPU}{Central Processing Unit}

\newacronym{SDU}{SDU}{Service Data Unit}
\newacronym{CBS}{CBS}{Code Block Size}
\newacronym{CB}{CB}{Code Block}
\newacronym{SPMD}{SPMD}{Single Program Multiple Data}
\newacronym{SIMD}{SIMD}{Single Instruction Multiple Data} 

\newacronym{SINR}{SINR}{Signal-to Interference Noise Ratio}
\newacronym{CO}{CO}{Central Office}
\newacronym{CA}{CA}{Carrier Aggregation}
\newacronym{SRS}{SRS}{Sound Reference Signal}
\newacronym{SC-OFDMA}{SC-OFDMA}{Single Carrier - Orthogonal Frequency Division Multiple Access}
\newacronym{FPGA}{FPGA}{Field-Programmable Gate Array}
\newacronym{TA}{TA}{Time Advancing}
\newacronym{CoMP}{CoMP}{Coordinated Multi-point}
\newacronym{NPRB}{NPRB}{Number of Physical Resource Blocks}
\newacronym{RTT}{RTT}{Round Trip Time}
\newacronym{CPRI}{CPRI}{Common Public Radio Interface}
\newacronym{CBR}{CBR}{Constant Bit Rate}
\newacronym{NRB}{NRB}{Number of Resource Blocks}
\newacronym{BJF}{BJF}{Biggest Job First}
\newacronym{EDF}{EDF}{Earliest Deadline First}
\newacronym{FCFS}{FCFS}{First-come, First-served}
\newacronym{PSTN}{PSTN}{Public Switched Telephone Network}
\newacronym{ETSI}{ETSI}{European Telecommunications Standards Institute}
\newacronym{vBBU}{vBBU}{virtualized BBU}
\newacronym{vRAN}{vRAN}{virtualized RAN}
\newacronym{IoT}{IoT}{Internet of Things}
\newacronym{B2B}{B2B}{Business to Business}
\newacronym{B2C}{B2C}{Business to Customer}
\newacronym{QoE}{QoE}{Quality of Experience}
\newacronym{QoS}{QoS}{Quality of Service}
\newacronym{VNO}{VNO}{Virtual mobile Network Operator}
\newacronym{SLA}{SLA}{Service Level Agreement}
\newacronym{VRRM}{VRRM}{Virtual Radio Resource Management}
\newacronym{KVM}{KVM}{Kernel-based Virtual Machine}
\newacronym{LXC}{LXC}{Linux Containers}
\newacronym{PS}{PS}{Processor Sharing}

\newacronym{eCPRI}{eCPRI}{evolved CPRI}
\newacronym{RoE}{RoE}{Radio over Ethernet}
\newacronym{PAPR}{PAPR}{Peak-to-average power ratio}
\newacronym{SC-FDMA}{SC-FDMA}{Single Carrier Frequency Division Multiple Access}
\newacronym{AGC}{AGC}{Automatic Gain Control}
\newacronym{PMD}{PMD}{Polarization Mode Dispersion}
\newacronym{ADC}{ADC}{Analogic-Digital Converter}

\newacronym{IQ}{IQ}{In-Phase Quadrature}
\newacronym{xRAN}{xRAN}{extensible Radio Access Network}
\newacronym{ISI}{ISI}{Inter-symbol interference}
\newacronym{FFT}{FFT}{Fast Fourier Transform}
\newacronym{IPC}{IPC}{Inter process communication}
\newacronym{CCDU}{CCDU}{Channel Coding Data Unit}
\newacronym{CC}{CC}{Channel Coding}
\newacronym{gNB}{gNB}{next-Generation Node B}
\newacronym{EUTRAN}{EUTRAN}{Evolved Universal Terrestrial Radio Access Network}
\newacronym{SCTP}{SCTP}{Stream Control Transmission Protocol}
\newacronym{NR}{NR}{New Radio}
\newacronym{NF}{NF}{Network Function}
\newacronym{CU}{CU}{Central Unit}
\newacronym{DU}{DU}{Distributed Unit}
\newacronym{NGC}{NGC}{Next Generation Core}
\newacronym{DL}{DL}{down-link}
\newacronym{UL}{UL}{up-link}
\newacronym{LJF}{LJF}{Largest Job First}
\newacronym{RANaaS}{RANaaS}{RAN as a Service}
\newacronym{NaaS}{NaaS}{Network as a Service}
\newacronym{NS}{NS}{Network Service}
\newacronym{FG}{FG}{Forwarding Graph}
\newacronym{VNFC}{VNFC}{VNF Component}
\newacronym{MANO}{MANO}{Management and Orchestration}
\newacronym{FIFO}{FIFO}{First In Firs Out}
\newacronym{NFVI}{NFVI}{NFV Infrastructure}
\newacronym{NFVO}{NFVO}{NFV Orchestrator}
\newacronym{PoP}{PoP}{Point of Presence}
\newacronym{NAT}{NAT}{Network Address Translation}
\newacronym{CDN}{CDN}{Content Delivery Network}
\newacronym{VNFM}{VNFM}{VNF Manager}
\newacronym{EM}{EM}{Element Management}
\newacronym{VIM}{VIM}{Virtualised Infrastructure Manager}

\newacronym{e2e}{e2e}{end-to-end}

\newacronym{AMF}{AMF}{Access and Mobility Management Function}
\newacronym{SMF}{SMF}{Session Management Function}
\newacronym{UPF}{UPF}{User Plane Function}
\newacronym{PCF}{PCF}{Policy Control Function}
\newacronym{UDM}{UDM}{Unified Data Management}
\newacronym{NRF}{NRF}{NF Repository Function}
\newacronym{AUSF}{AUSF}{Authentication Server Function}
\newacronym{API}{API}{Application Programming Interface}
\newacronym{HSS}{HSS}{Home Subscriber Server}
\newacronym{PCRF}{PCRF}{Policy and Charging Rules Function}
\newacronym{SOA}{SOA}{Software-Oriented Architecture}
\newacronym{AKA}{AKA}{Authentication and Key Agreement}
\newacronym{AF}{AF}{Application Function}
\newacronym{NEF}{NEF}{Network Exposure Function}
\newacronym{NSSF}{NSSF}{Network Slice Selection Function}
\newacronym{NSSP}{NSSP}{Network Slice Service Profile}
\newacronym{VES}{VES}{Virtual Event Streaming}
\newacronym{NSSAI}{NSSAI}{Network Slice Selection Assistance Information}

\newacronym{NSSI}{NSSI}{Network Slice Subnet Instance}

\newacronym{NSS}{NSS}{Network Slice Subnet}
\newacronym{NSC}{NSC}{Network Slice Customer}
\newacronym{NSP}{NSP}{Network Slice Provider}
\newacronym{CSC}{CSC}{Communication Service Customer}
\newacronym{CSP}{CSP}{Communication Service Provider}
\newacronym{SST}{SST}{Slice/Service Type}
\newacronym{SD}{SD}{Slice Differentiator}
\newacronym{USRP}{USRP}{UE Router Selection Policy}
\newacronym{S-NSSAI}{S-NSSAI}{Single Network Slice Selection Assistance Information}
\newacronym{ONISTT}{ONISTT}{Open Net-centric Interoperability Standards for Training and Testing}
\newacronym{KB}{KB}{Knowledge Base}
\newacronym{NSI}{NSI}{Network Slice Instance}

\newacronym{VF}{VF}{Virtual Function}
\newacronym{VFC}{VFC}{Virtual Function Component}
\newacronym{CR}{CR}{Complex Resource}
\newacronym{PNF}{PNF}{Physical Network Function}
\newacronym{CP}{CP}{Connection Point}
\newacronym{VL}{VL}{Virtual Link}
\newacronym{SDC}{SDC}{Service Design and Creation}
\newacronym{ONAP}{ONAP}{Open Network Automation Platform}
\newacronym{VID}{VID}{Virtual Infrastructure Deployment}
\newacronym{VSP}{VSP}{Vendor Software Product}

\newacronym{WEF}{WEF}{Wireless Edge Factory}

\newacronym{DP}{DP}{Data Plane}
\newacronym{ECOMP}{ECOMP}{Enhanced Control Orchestration Management and Policy}

\newacronym{AAI}{AAI}{Active and Available Inventory}
\newacronym{SDNC}{SDNC}{Software Defined Network Controller}
\newacronym{SO}{SO}{Service Orchestrator}
\newacronym{APPC}{APPC}{Application Controller}
\newacronym{DCAE}{DCAE}{Data Collection Analytics and Events}
\newacronym{OOF}{OOF}{ONAP Optimization Framework}
\newacronym{OSS}{OSS}{Operation Support System}
\newacronym{BSS}{BSS}{Business Support System}
\newacronym{SOCKS}{SOCKS}{Secured Over Credential-based Keberos}
\newacronym{VVP}{VVP}{VNF Validation Program}
\newacronym{PDP}{PDP}{Policy Decision Point}
\newacronym{PEP}{PEP}{Policy Enforcement Point}
\newacronym{PCC}{PCC}{Policy Creation Component}
\newacronym{RU}{RU}{Remote Unit}
\newacronym{ORAN}{ORAN}{Open-RAN}
\newacronym{CFS}{CFS}{Customer Facing Service}
\newacronym{RFS}{RFS}{Resource Facing Service}

\newacronym{VLM}{VLM}{Vendor License Model}
\newacronym{OCN}{OCN}{Open Core Network}
\newacronym{NBI}{NBI}{Northbound Interface}
\newacronym{I/Q}{I/Q}{in-phase / in-quadrature}
\newacronym{CUPS}{CUPS}{Control User Plane Separation}